\documentclass{article}

\usepackage{arxiv}

\usepackage[utf8]{inputenc} 
\usepackage[T1]{fontenc}    
\usepackage{url}            
\usepackage{booktabs}       
\usepackage{amsfonts}       
\usepackage{nicefrac}       
\usepackage{microtype}      
\usepackage{lipsum}
\usepackage{graphicx}
\usepackage{natbib}
\usepackage{float}
\usepackage{multirow}
\usepackage{array}
\usepackage{subcaption}
\usepackage{adjustbox}
\usepackage{setspace}
\usepackage{makecell}
\usepackage[table]{xcolor}   
\usepackage{siunitx}  
\usepackage{longtable}
\usepackage{hyperref}       

\graphicspath{ {./images/} }

\newcommand{\RedText}[2]{\textcolor{red!#1}{#2}}

\newcommand{\SeverityText}[3]{%
  \ifnum#1=0 0%
  \else
    \ifnum#2=0 \RedText{#3}{#1}
    \else
      \ifnum#2=1 \RedText{#3}{#1}
      \else \RedText{#3}{#1}
      \fi
    \fi
  \fi
}

\title{Protecting Young Users on Social Media: Evaluating the Effectiveness of Content Moderation and Legal Safeguards on Video Sharing Platforms}

\author{
 Fatmaelzahraa Eltaher \\
  School of Computer Science\\
  Technological University Dublin\\
  Dublin\\
  \texttt{fatmaelzahraa.eltaher@tudublin.ie} \\
   \And
 Rahul Krishna Gajula \\
  School of Computer Science\\
  Technological University Dublin\\
  Dublin \\
  \texttt{rahul.gajula@tudublin.ie} \\
  \And
 Luis Miralles-Pechuán \\
  School of Computer Science\\
  Technological University Dublin\\
  Dublin \\
  \texttt{luis.miralles@tudublin.ie} \\
  \And
 Patrick Crotty \\
  School of Medicine\\
  Trinity College Dublin\\
  Dublin \\
  \texttt{pcrotty@tcd.ie} \\
  \And
 Juan Martínez-Otero \\
  School of Law\\
  University of Valencia\\
  Valencia \\
  \texttt{juan.maria.martinez@uv.es} \\
  \And
 Christina Thorpe \\
  School of Informatics and Cybersecurity\\
  Technological University Dublin\\
  Dublin \\
  \texttt{christina.thorpe@tudublin.ie} \\
  \And
 Susan McKeever \\
  School of Computer Science\\
  Technological University Dublin\\
  Dublin \\
  \texttt{susan.mckeever@tudublin.ie} \\
}

\begin{document}
\maketitle
\begin{abstract}
Video-sharing social media platforms, such as TikTok, YouTube, and Instagram, implement content moderation policies aimed at reducing exposure to harmful videos among minor users. As video has become the dominant and most immersive form of online content, understanding how effectively this medium is moderated for younger audiences is urgent. In this study, we evaluated the effectiveness of video moderation for different age groups on three of the main video-sharing platforms: TikTok, YouTube, and Instagram. We created
experimental accounts for the children assigned ages 13 and 18. Using these accounts, we evaluated 3,000 videos served up by the social media platforms, in passive scrolling and search modes, recording the frequency and speed at which harmful videos were encountered. Each video was manually assessed for level and type of harm, using definitions from a unified framework of harmful content. 

The results show that for passive scrolling or search-based scrolling, accounts assigned to the age 13 group encountered videos that were deemed harmful, more frequently and quickly than those assigned to the age 18 group. On YouTube, 15\% of recommended videos to 13-year-old accounts during passive scrolling were assessed as harmful, compared to 8.17\% for 18-year-old accounts. On YouTube, videos labelled as harmful appeared within an average of 3:06 minutes of passive scrolling for the younger age group. Exposure occurred without user-initiated searches, indicating weaknesses in the algorithmic filtering systems. These findings point to significant gaps in current video moderation practices by social media platforms. Furthermore, the ease with which underage users can misrepresent their age demonstrates the urgent need for more robust verification methods.
\end{abstract}


\section{Introduction}
Social media has changed how most people connect, communicate, and share content, reaching over 5 billion users in 2024 and projected to surpass 6 billion by 2028 \citep{socialMediaUsers}. Leading platforms such as Facebook, YouTube, Instagram, TikTok, and Snapchat collectively engage billions of monthly users, making social networking one of the most popular online activities \citep{popularSocialNetworks}.

\begin{table}
    \centering
    \caption{Most popular social networks worldwide as of April 2024, by number of monthly active users (in millions). Source: \citep{popularSocialNetworks}}.
    \label{tab:active_users}
    \begin{tabular}{l r}
        \hline
        \textbf{Platform} & \textbf{Active Users} \\
        \hline
        Facebook            & 3,065 M \\
        YouTube             & 2,504 M \\
        Instagram           & 2,000 M \\
        WhatsApp            & 2,000 M \\
        TikTok              & 1,582 M\\
        WeChat              & 1,343 M\\
        Facebook Messenger  & 1,010 M\\
        Telegram            & 900 M\\
        Snapchat            & 800 M  \\
        Douyin              & 755 M  \\
        Kuaishou            & 700 M  \\
        X/Twitter           & 611 M  \\
        Weibo               & 598 M  \\
        QQ                  & 554 M  \\
        Pinterest           & 498 M  \\
        \hline
    \end{tabular}
\end{table}

As shown in Table \ref{tab:active_users}, Facebook and YouTube dominate global user engagement. In England, 91\% of teenagers aged 13 to 18 and 65\% of children aged 8 to 12 actively participate on social media \citep{Digital_childhoods}. In the United States, approximately 60\% of teens aged 13 to 17 frequently use platforms including Instagram, Snapchat, and TikTok \citep{anderson2023teens}, while even younger children regularly access platforms such as YouTube and TikTok usually with the knowledge and consent of their parents \citep{KKSO, WeProtectGlobalAlliance, Childwise}.

A significant portion of this engagement is now driven by video content, which has become the dominant and most immersive medium on social platforms. The rise of short-form video, popularised by TikTok and quickly mirrored by platforms like Instagram Reels and YouTube Shorts, represents a major shift in online engagement. These easily consumable, rapidly delivered videos are especially appealing to younger audiences, who increasingly prefer quick, visually engaging content over text or static images \citep{10.1145/3614419.3644023}. This shift has led to a notable increase in social media use among children and teenagers, many of whom now use platforms like TikTok and Snapchat as their primary sources of entertainment, social interaction, and even news \citep{LIU2024108007,Ge18082021}

Social media platforms use algorithms to personalise content recommendations based on user interests to keep users engaged. While personalisation enhances user experiences, it has sparked significant concerns regarding the amplification of potentially harmful content \citep{gorwa2020algorithmic}. UNICEF defines harmful content as anything—be it an image, video, or text—that offends, upsets, or causes harm to individuals \citep{unicef}. To address these concerns, platforms deploy various moderation strategies, ranging from simple content warnings to content removal or suspension of user accounts \citep{fiesler2018reddit,seering2019moderator,gillespie2018custodians}.

Content moderation generally falls into three approaches: volunteer moderation in smaller online communities, commercial moderation on larger platforms, and automated moderation powered by artificial intelligence \citep{gillespie2018custodians}. Given the enormous volume of daily content, AI-driven moderation systems have become increasingly effective for detecting and filtering inappropriate material \citep{chandrasekharan2019crossmod}. Furthermore, on January 7, 2025, Meta announced substantial revisions to Facebook and Instagram's content moderation procedures. Meta intends to terminate its fact-checking program in the United States and implement a community-driven approach similar to X's Community Notes feature \citep{factchecking}.

Platforms typically classify content into three distinct categories: (1) permitted content, complying fully with community guidelines; (2) prohibited content, such as hate speech, violent threats, or harassment, which must be removed immediately; and (3) restricted content, which remains accessible under specific conditions, such as through age verification processes or explicit warning screens \citep{Facebook_Community_Standards, Instagram_Community_Standards, Instagram_warning_screen, AgeRestrictionsYouTube, YouTube_Community_Standards, TikTok_content_moderation,apply_age_restrictions_Twitter}.

In response to growing concerns regarding child safety, platforms have introduced age-based restrictions to limit exposure to potentially harmful content. TikTok, for instance, provides a parental-controlled ``Restricted Mode" and automatically restricts access to certain age-sensitive content for users between 13 and 17 \citep{Restricted_Mode_on_TikTok,age_restricted_content_on_TikTok_LIVE}. Similarly, YouTube employs machine learning to classify age-restricted videos and encourages content creators to label their uploads correctly \citep{efforts_to_protect_YouTube, AgeRestrictionsYouTube}. Facebook, Instagram, and Twitter have implemented similar restrictions designed to protect minors \citep{efforts_to_protect_Facebook,apply_age_restrictions_Twitter}.
However, social platforms primarily rely on \textit{self-declared} age at account setup time, making it easy for minors to fake their age or bypass controls. This absence of robust age verification raises significant doubts about the effectiveness or reach of age-related controls. Previous works indicate that content moderation alone cannot effectively safeguard minors unless supported by strict, verifiable age verification mechanisms \citep{icissp25}.

Despite these safeguards, evidence shows that children encounter harmful content online. A survey by the Children's Commissioner for England indicated that nearly 45\% of children aged 8 to 17 have encountered harmful content on social media \citep{Digital_childhoods}. Moreover, a study across 25 European countries revealed that 20\% of children aged 9 to 16 were exposed to sexual content online \citep{staksrud2013does}. Further research highlights algorithmic risks, demonstrating that toddlers watching child-friendly videos on YouTube have a 3.5\% chance of encountering inappropriate content within ten recommended videos \citep{papadamou2020disturbed}. 

As video becomes the primary mode of engagement for audiences, it is essential to assess how effectively video-sharing platforms safeguard younger audiences from harmful videos and to examine whether age-based moderation and legal frameworks reduce the risks minors face. As covered in Section \ref{sec:Literature_Review}, previous research often focuses on specific harmful categories (e.g., misogynistic, mental health issues), with less attention to comprehensive analyses across multiple platforms and varying interaction modes (passive vs. search-based). Our work addresses this gap by systematically evaluating content moderation in TikTok, YouTube, and Instagram, offering a cross-platform comparison of moderation for minors versus adults. The specific research questions addressed in this paper are:

\begin{itemize}
    \item How does the effectiveness of content moderation on leading video-sharing platforms differ for minors compared to adults?
    \item How quickly do harmful videos surface when user behaviour varies from passive scrolling of videos to active searching?
    \item What types of harmful videos are most commonly encountered by users across different platforms and interaction modes, and how do these content types vary in severity?
    \item To what extent do the content moderation guidelines of video-sharing platforms align with their actual enforcement practices?
\end{itemize}

The remainder of the paper is structured as follows. Section \ref{sec:Literature_Review} explores existing research on content moderation practices and regulatory frameworks. Section \ref{sec:Methodology} details our approach and experimental setup for evaluating the effectiveness of moderation. Section \ref{sec:Analysis_of_the_Results}  discusses the empirical outcomes regarding age-based exposure and user engagement. Section \ref{sec:Conclusion} highlights implications, provides practical recommendations for platform improvements, and outlines avenues for future research.

\section{Literature Review} \label{sec:Literature_Review}

This section reviews the current literature on the limitations of content moderation, focusing on protecting minors from harmful online content. It begins by analysing empirical studies that reveal the shortcomings of existing moderation practices on popular social media platforms. This is followed by an overview of the legal frameworks in the United States, the European Union, and the United Kingdom that attempt to address these risks.
\subsection{Content Moderation in Video Sharing Platforms}

Effective content moderation is critical in safeguarding minors online, particularly those who bypass age verification mechanisms. Between October and December 2024, TikTok removed over 153 million videos for policy violations, while YouTube took down nearly 9.5 million. Despite platform efforts, research consistently shows that children can encounter harmful material through passive exposure and active interaction \citep{papadamou2020disturbed,fibrilla2021exposure}. This has prompted empirical investigations into how social media algorithms amplify potentially damaging content.

Some studies examined how passively watching suggested videos, without any interaction, can lead users to harmful content. A study by Amnesty International focused on the
prevalence of mental health content on TikTok \citep{amnestyinternational2023darkness}. Forty accounts were created, all simulating 13-year-old users. Within 5 to 6 hours of use, nearly half of the videos shown to accounts expressing interest in
mental health were classified as potentially harmful.

Similarly, Regehr et al. \citep{regehr2024safer} conducted an experiment on TikTok to assess how easily users can access misogynistic and manosphere content. The findings revealed that the likelihood of encountering misogynistic content increased fourfold within five days, illustrating how rapidly ideologically charged material can become embedded in a user’s feed.

Other studies examine the effect of engagement actions, such as liking and searching, on exposing users to certain types of dangerous videos. Baker et al. \citep{baker2024recommending} analysed ten user accounts on YouTube Shorts and TikTok to explore the online experiences of boys and young men. They focused on exposure to problematic content, including manosphere ideologies, anti-feminist messages, racism, and anti-LGBTQ sentiments. Their findings revealed that YouTube Shorts featured a significantly higher proportion of such content (61.5\%) compared to TikTok (34.7\%). 

In a related experiment, Williams et al. \citep{williams2021surveilling} created five TikTok accounts simulating 13-year-old girls. Researchers aimed to assess how quickly the platform's algorithm recommended questionable content, including ethnic or gender stereotypes, misinformation about COVID-19 or vaccines, or weight loss content. After viewing 600 videos, they found that one in four promoted harmful ethnic stereotypes and one in five reinforced damaging gender stereotypes. However, no weight-loss or COVID-19 misinformation was shown, possibly due to TikTok’s content moderation efforts or limitations in the study design.

An Australian study \citep{australia2022algorithms} investigated how YouTube and YouTube Shorts expose boys and young men to misogynistic content. Users were progressively recommended more radical content by liking and following mainstream and extreme influencers. YouTube Shorts, in particular, escalated the delivery of misogynistic and incel-related material.

Similarly, the Centre for Countering Digital Hate \citep{DeadlybyDesign} reported that TikTok served suicide-related content within 2.6 minutes and eating disorder content within 8 minutes of use by simulated teen accounts. Every 39 seconds, TikTok displays content about teens’ mental health and body image.

Ekō, an international advocacy organisation \citep{Eko}, investigated harmful content on TikTok using an account registered to a 13-year-old user \citep{TikTok’sdeadlyalgorithm}. Their focus included suicide, incel ideology, the manosphere and drug-related content. The study concluded that a mere ten minutes of limited engagement with suicide-oriented content prompted TikTok’s algorithm to push further self-harm and suicide-related material to these underage users.

Papadamou et al. \citep{papadamou2020disturbed} explored YouTube’s recommendation mechanisms. The study found that even when starting from safe content, there was a 3.5\% chance of reaching inappropriate material within ten recommendations.

\subsubsection{Gaps in Current Research and the Present Study's Contribution:}

These studies highlight gaps in moderation policies that permit harmful videos to be served to young or otherwise vulnerable audiences. While existing research often examines TikTok or YouTube about specific harm categories, such as misogynistic narratives or mental health content, there is limited investigation into the full breadth of harmful videos. Accurate moderation hinges on well-defined categories of harm; yet key areas such as Child Sexual Abuse Material (CSAM), ``Sensitive and Mature Themes’’ and spam or fraud remain underexplored \citep{metaStandards,tiktokCommunityGuidelines,YouTubeCommunityGuidelines}.

In addition, Instagram—one of the most widely used social networks—has been largely absent from prior examinations, leaving significant gaps in our understanding of its content moderation efficacy. To address these deficits, this work evaluates video moderation practices on multiple video-sharing platforms, including YouTube, Instagram, and TikTok, and captures a broader range of harmful content categories. Moreover, it investigates how searching and scrolling behaviours influence the discovery of such material, offering a more comprehensive view of harmful content distribution across platforms.

\subsection{Legislation on content moderation}

Considering the importance that digital platforms have gained from a political (as a new public space), cultural (shaping the collective imagination), and social (as the primary source of entertainment for millions of people) standpoint, public authorities have progressively taken charge of their regulation. Although content moderation regulation was initially left to the platforms themselves, more recently, in most jurisdictions, various legal instruments have started to address this matter, such as the Digital Services Act (DSA), 2022 in the EU and the Online Safety Act, 2023 in the  UK, and the Communications Decency Act in the US.

Moderation carried out by social networks, which can manifest in the deletion of the content, the reduction of items' visibility or the suspension of users, no longer has an exclusively private scope but also carries a significant public dimension, affecting individuals' freedom of expression and the formation of a free public opinion. R. Rexhepi \citep{rexhepi2023content} wrote, “Social media networks have become forums and mediums of important conversation, and the responsibility to regulate it is too great for platforms to be left to deal with it alone.”  

We provide a brief overview of existing laws in the United States, the European Union (EU), and the United Kingdom to offer a broad perspective on regulating content moderation on social media.

\subsubsection{United States}

At the federal level, the United States lacks specific regulations governing content moderation practices and policies of online platforms.

The First Amendment and Section 230 of the Communications Decency Act determine the key regulatory framework in content moderation. Section 230 shields interactive computer service providers, including social media platforms, and their users from liability for publishing - and, in certain instances, restricting access to or availability of - content posted by others. The purpose of Section 230 is to foster free speech while allowing interactive computer service providers to moderate content without government interference.

In alignment with the First Amendment, Section 230 establishes a notably hands-off regulatory framework, granting platforms significant autonomy and broad immunity in their content moderation practices \citep{gill2020regulating}. This legal approach has been subject to two main criticisms: first, the wide discretion it affords providers in moderating and removing content, which can potentially enable covert forms of censorship; second, the insufficient protections it offers to citizens against harmful content, particularly concerning the protection of minors from inappropriate material.

Several regulatory proposals have recently been introduced in response to growing concerns about content moderation, aiming to explicitly encourage, discourage, prohibit, or mandate specific moderation practices on digital platforms \citep{cho2025social}. Although its final approval remains uncertain, one of the most notable legislative proposals is the KOSA (H.R.7891 - Kids Online Safety Act, 118th Congress, 2023-2024). Its provisions on the protection of minors and content moderation (Sections 4 and 6) are similar to those currently in force in the EU and the United Kingdom. In summary, KOSA aims to establish a protective and preventive framework that ensures content moderation adheres to standards of transparency, accountability, and procedural fairness.

Whether this regulation will be enacted or US authorities will pursue alternative public policies regarding content moderation remains unclear. Among the various proposals under consideration are: (1) maintaining the current hands-off regime; (2) encouraging changes to moderation systems through hearings or investigations; (3) regulating moderation to require specific conditions regarding speed and transparency, as KOSA proposes; (4) implementing federal advisory or regulatory oversight over social media with certain powers over platforms; and (5) pursuing alternative measures such as digital education initiatives \citep{cho2025social}.

\subsubsection{European Union}

Within the EU, two principal regulatory frameworks govern content moderation on digital platforms. Both frameworks impose a duty of care on platforms, requiring them to take reasonable steps to ensure user safety and to address illegal and harmful activities.

The first framework stems from the Audiovisual Media Services Directive (AVMSD) 2018, which has been transposed into national legislation across Member States. It applies specifically to video-sharing platforms (VSPs) based in the EU, as well as to prominent users such as influencers. Article 28b explicitly refers to video-sharing platforms. Paragraph 1 mandates measures to prevent minors from accessing harmful content and to prevent the public from accessing illegal content (e.g. hate speech). Article 28b, paragraph 3, outlines various potential measures that Member States can impose on VSPs—many of which pertain to content moderation mechanisms. These measures include establishing systems for rating and reporting content, as well as processes for handling user complaints. 

The second key regulation is the Digital Services Act of 2022 \citep{union2022regulation}. This regulation applies to digital platforms that offer services to citizens in EU Member States, including significant platforms such as X, Instagram, YouTube, and many adult content websites. It imposes specific proactive obligations on providers to protect minors from potentially harmful content \footnote{See Recitals 71 and 89 RSD. Regarding large platforms, Recital 89 states they ``should take measures to protect minors from content that may harm their physical, mental or moral development and provide tools that allow conditional access to such information".}.

Concerning content moderation, the Digital Services Act aims to enhance transparency in content moderation systems by requiring platforms to publish moderation policies (Article 14), annual reports (Article 15), and the rationale behind their decisions (Articles 16-17). It also strengthens procedural due process guarantees by ensuring the right to appeal and the involvement of human agents in the decision-making process (Arts. 20, 21). Furthermore, large platforms and search engines must conduct impact assessments to identify risks to minors within their environments, including those posed by platform algorithms, and implement measures to mitigate them (Arts. 34, 35). 

To summarise, the EU legal framework aims to guarantee that content moderation systems are fair, transparent and effective, contributing to a safer digital environment for all users, particularly minors. However, recent research suggests that many platforms are not fully meeting these obligations in practice, particularly regarding transparency and algorithmic risk assessments. In any case, when referring to content moderation, the regulation focuses on reporting and removing content rather than its ``algorithmic” management, which may foster or reduce its visibility among users \citep{gillespie2022not}.

\subsubsection{United Kingdom}

In the United Kingdom, the governing legislation is the Online Safety Act 2023, which imposes a series of obligations on online service providers to enhance the protection of their users.
The Act sets forth a framework of relatively broad provisions and guiding principles, which will be further refined through secondary legislation (see Sections 12, 29, 30). Like EU regulations, the UK framework establishes general obligations and specific duties applicable to more prominent social media platforms. 
Regarding the protection of minors and content moderation, the Online Safety Act 2023 contains provisions similar to those enacted in the EU. A brief overview of the most significant provisions is outlined below:

\begin{itemize}
    \item Sections 11 and 12 address the protection of minors online, mandating that platforms conduct risk assessments and implement monitoring mechanisms to mitigate potential harms.
    
    \item Sections 20, 21, and 22 govern the systems for lodging and resolving complaints while also imposing a duty to uphold freedom of expression within social media environments. Notably, these sections require platforms to incorporate clear information regarding the policies and procedures for handling and resolving relevant complaints into their terms of service, accessible to all users, including minors.
    
    \item Sections 71 and 72 regulate the processes by which users may contest actions taken against them or their content by the platform. Such procedures must be explicitly outlined within the terms of use.
    
    \item Section 77 obligates larger platforms to publish an annual transparency report containing detailed information on various operational aspects, including content moderation practices.
\end{itemize}

The provisions of the Online Safety Act are rather broad and must, therefore, be further specified by the UK’s audiovisual authority, Ofcom. In one of its initial public consultations aimed at developing the Online Safety Act, Ofcom \citep{Ofcom2024} sets out ten specific measures regarding content moderation, underscoring the importance it intends to place on such protective techniques.

Although the British regulation is still in its initial phase of implementation, it is apparent that, in matters of content moderation, it aligns with the provisions outlined in the EU Digital Services Act: transparency in terms of use and specific moderation processes; active disclosure of annual actions in this area; and procedural due process safeguards—such as the right to appeal—for affected users. As we noted regarding European regulations on content moderation, UK legislation focuses more on reporting, assessing, and removing content than on platform algorithmic recommendations.

To conclude, analysing the legal frameworks in the United States, the European Union, and the United Kingdom reveals two key trends. First, the rise of regulatory initiatives—such as the EU’s Digital Services Act (DSA), the UK’s Online Safety Act, and the proposed Kids Online Safety Act (KOSA) in the US—reflects growing public and political pressure to oversee content moderation practices. In the EU, this has already translated into concrete enforcement actions. In 2024, the European Commission launched formal investigations into platforms such as X, TikTok, AliExpress, Facebook, Instagram, and Temu for potential DSA violations \citep{fabbri2025role}. Meanwhile, the US continues to debate whether and how to legislate in this area.

Second, as of 2025, regulation primarily targets transparency and accountability in reporting and content removal. The DSA requires Very Large Online Platforms (VLOPs) with over 45 million EU users to publish annual risk assessment reports addressing systemic risks like disinformation and harm to minors. Starting 17 February 2024, all intermediary service providers must release yearly transparency reports on their content moderation practices. Platforms must also disclose their moderation policies, provide clear complaint mechanisms, and ensure users have a fair due process. However, while the DSA enhances transparency in algorithmic recommendation systems, specific strong regulations to protect minors from harmful content via these systems remain limited. 

While these frameworks mark significant progress, a regulatory focus on post-exposure responses may be insufficient. Future efforts must more directly tackle the algorithmic amplification of harmful content to protect young users meaningfully.

\section{Study Methodology}  \label{sec:Methodology}
\begin{figure}
\centerline{%
\includegraphics[width=0.35\textwidth]{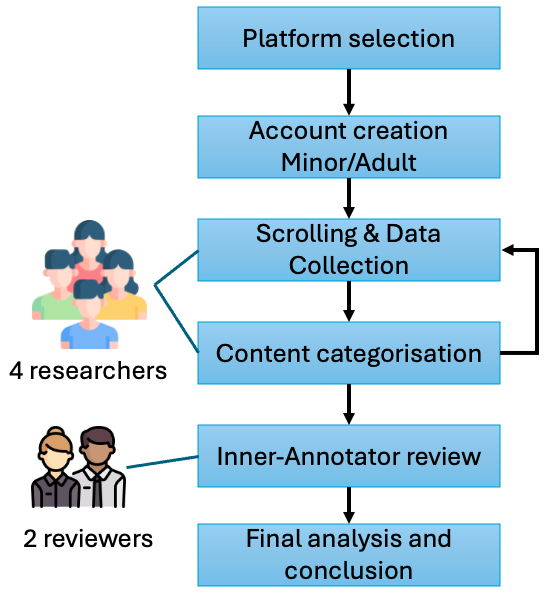}%
}%
\caption{High-level methodology diagram illustrating the experimental sequence, including account creation and labelling.}
\label{fig:simple_methodology}
\end{figure}

Our study examines how effectively social media video sharing platforms protect minors from harmful or inappropriate content.  Our objectives are to: 
\begin{enumerate}
    \item Compare the efficacy of content moderation systems for minors versus adults on leading video sharing platforms.
    \item Measure the speed at which harmful material surfaces under different interaction styles (passive scrolling versus explicit searching).
    \item Assess whether the community guidelines of each platform align with their real-world enforcement practices.
\end{enumerate}
Four researchers took part in video annotation separately. Each researcher managed a distinct set of accounts under standardised conditions; subsequently, two additional experts reviewed the labelled content to verify inter-annotator agreement. Figure~\ref{fig:simple_methodology} offers a visual overview of our experimental flow, from platform selection through final data analysis.

\subsection{Platform Selection}

To capture highly relevant, mainstream user experiences, we focused on social media platforms where short-form videos and infinite scrolling feeds are the primary ways content is delivered. Two main criteria guided this selection: 
\begin{enumerate}
    \item The platform must provide algorithm-driven endless scrolling, continuously exposing users to recommended videos.
    \item The platform must have a sufficiently large user base to represent broader social media usage patterns.
\end{enumerate}

Based on these two points and the most popular social media platforms as seen in Table \ref{tab:active_users}, YouTube, Instagram, and TikTok were chosen due to their large user bases and robust, continuous-scrolling mechanisms for short videos. Consequently, our study focuses on Instagram Reels, TikTok’s "For You" feed, and YouTube Shorts.

\subsection{Creation of a Unified Framework for Harmful Content}

This study aims to identify the types of harmful content present on the platforms. Therefore, we need a list of harmful content categories, definitions and examples that act as a guideline.

Each selected platform operates under its own Community Guidelines \citep{metaStandards,tiktokCommunityGuidelines,YouTubeCommunityGuidelines}, outlining permissible content and enforcement measures. These policies, however, vary considerably. To conduct a consistent cross-platform evaluation, we established a \emph{Unified Harmful Content Framework} that reconciles divergent content categorisations into a single taxonomy.

We began by reviewing each platform’s policy documents, capturing every category of harmful content. Categories present on one platform but absent on others were integrated into a single list, thus ensuring broad coverage. For example, Meta explicitly mentions ``Privacy Violations,'' whereas TikTok’s guidelines highlight ``Dangerous Challenges and Activities'' as a distinct category. Incorporating each platform’s unique categories yielded the framework \footnote{This comprehensive taxonomy is publicly accessible: \url{https://tudublin.sharepoint.com/:x:/r/sites/N-LightTechCoalition/Shared\%20Documents/General/Content\%20Moderation\%20Research/Content\%20Moderation\%20Policies/Social\%20Media\%20Platforms\%20Community\%20Guidelines\%20New.xlsx?d=we4db6e5e95a948118138e8b840602c28\&csf=1\&web=1\&e=ZFf4Ii}} summarized in Table~\ref{tab:content-categories}.

\begin{table}
    \small\sf\centering
    \caption{Summary of Content Categories\label{tab:content-categories}}
    \begin{adjustbox}{max width=\textwidth}
    \begin{tabular}{lp{4cm}p{7.5cm}}
    \toprule
    \textbf{Category} & \textbf{Subcategory} & \textbf{Definition} \\
    \midrule
    \multirow{4}{*}{\textbf{Violent and Criminal Behaviour}} 
      & Coordinating Harm & Organising or inciting harmful actions, including real-world and digital crimes. \\ 
      & Dangerous Organisations & Groups promoting violence, hate, or large-scale harm. \\ 
      & Violence and Incitement & Calls for violence, threats, or glorification of harmful acts. \\ 
      & Criminal Organisations & Large-scale criminal groups such as cartels or terrorist networks. \\ 
    \midrule
    \multirow{4}{*}{\textbf{Hate Speech and Hateful Behaviour}} 
      & Hate Speech & Statements inciting hatred based on protected attributes. \\ 
      & Hateful Groups & Organisations promoting hate-based ideologies. \\ 
      & Cyberbullying & Repeated online harassment targeting individuals. \\ 
      & Harassment & Intimidation, doxxing, or persistent unwanted behaviour. \\ 
    \midrule
    \multirow{3}{*}{\textbf{Abuse and Exploitation}} 
      & Sexual and Physical Abuse & Non-consensual acts, coercion, or intimate violence. \\ 
      & Human Exploitation & Trafficking, forced labour, or illegal adoption. \\ 
      & Privacy Violations & Unauthorised sharing of sensitive personal data. \\ 
    \midrule
    \multirow{3}{*}{\textbf{Mental and Behavioural Health}} 
      & Self-Harm & Content encouraging self-harm or suicide. \\ 
      & Eating Disorders & Promoting unhealthy eating habits. \\ 
      & Dangerous Challenges & Urging participation in hazardous activities. \\ 
    \midrule
    \multirow{4}{*}{\textbf{Sensitive and Mature Themes}} 
      & Nudity & Regulation of sexual or explicit content. \\ 
      & Graphic Content & Depictions of violence, mutilation, or harm. \\ 
      & Sexual Services & Promotion or facilitation of sex work. \\ 
      & Animal Abuse & Cruel or harmful treatment of animals. \\ 
    \midrule
    \multirow{5}{*}{\textbf{Dis/Misinformation}} 
      & General Misinformation & False or misleading content undermining public trust. \\ 
      & Election Misinformation & Inaccurate claims compromising democratic processes. \\ 
      & Health Misinformation & Content contradicting scientific or medical consensus. \\ 
      & AI-Generated Media & Manipulated content such as deepfakes. \\ 
      & Conspiracy Theories & Unsubstantiated allegations targeting institutions or individuals. \\ 
    \midrule
    \multirow{2}{*}{\textbf{Regulated Goods}} 
      & Illegal Sales & Promotion, or trade of prohibited items (drugs, firearms). \\ 
      & Gambling & Promotion of gambling, alcohol, or drugs. \\ 
    \midrule
    \multirow{2}{*}{\textbf{CSAM}} 
      & Child Exploitation & Content sexualising or endangering minors. \\ 
      & Grooming & Deceptive interactions aimed at minors. \\ 
    \midrule
    \multirow{3}{*}{\textbf{Spam and Fraud}} 
      & Fake Engagement & Artificially boosting likes, views, or follows. \\ 
      & Impersonation & Misleading accounts mimicking real users or brands. \\ 
      & Fraud & Scams or deceptive financial schemes. \\ 
    \midrule
    \multirow{2}{*}{\textbf{Privacy and Security}} 
      & Personal Data & Sharing sensitive data, leading to risks. \\ 
      & Cybersecurity & Breaches or unauthorised access to systems. \\ 
    \midrule
    \multirow{3}{*}{\textbf{Legal Issues}} 
      & IP Violations & Copyright or trademark infringement. \\ 
      & Account Integrity & Measures preventing repeated policy violations. \\ 
      & Authentic Identity & Ensuring users represent real individuals. \\ 
    \midrule
    \multirow{3}{*}{\textbf{Enforcement Actions}} 
      & Protection of Minors & Removal of content harmful to children. \\ 
      & Local Laws & Compliance with jurisdiction-specific regulations. \\ 
      & User Requests & Account removals upon user request. \\ 
    \bottomrule
    \end{tabular}
    \end{adjustbox}
\end{table}

Differences between how platforms moderate content for minors versus adults are readily observable in their published Community Guidelines, particularly regarding explicit or violent material. For example, YouTube \citep{YouTube_CSAM_Policy} and Instagram \citep{Instagram_CSAM_Policy} specify a blanket prohibition on any sexual content featuring minors. Conversely, adult content deemed acceptable, such as artistic depictions, educational resources, or breastfeeding imagery, may be age-restricted or labelled rather than removed entirely. TikTok \citep{Tiktok_CSAM_Policy} similarly applies tighter filters to minor accounts, restricting violent or sexual content, which, for adult users, may carry only warnings or age gates. Collectively, these policies highlight a shared commitment to age-sensitive moderation strategies that aim to shield younger audiences.

To further outline the spectrum of harmful content, our study builds upon each platform’s enforcement policies \citep{metaStandards,MetaTransparencyCM,tiktokCommunityGuidelines,TikTok_content_moderation,YouTubeCommunityGuidelines,YouTubeTransparencyRemovals}, assigning severity levels to capture varying levels of risk. Low-severity material includes mild stereotypes or insensitive humour, while high-severity examples comprise direct violence or explicit hate speech. 

\subsection{User Interaction Modes} \label{User interaction modes}

We examined two distinct user interaction modes — \emph{passive} and \emph{search-based} — to determine how different behaviours influence the probability of encountering harmful content. By comparing these modes, we aimed to determine whether user intent, such as purposefully searching for specific topics, affects exposure rates, thereby informing content moderation strategies tailored to diverse usage patterns. Each experiment takes about 90-120 minutes on average.

\paragraph{Account 1: Passive Scrolling}
In this scenario, an account scrolls through 300 videos consecutively, each viewed for approximately 20 seconds, without performing additional actions like searches, likes, comments, or shares. Any non-English videos are immediately skipped to simplify the annotation process. This setup approximates the experience of a new, largely passive user, offering insight into how recommendation algorithms push potentially harmful material to individuals with minimal engagement.

\paragraph{Account 2: Search-Based Scrolling}
In contrast, search-based scrolling adopts a \emph{multi-stage} approach. First, the account views 100 videos passively, watching each for 20 seconds. It then executes a search using ``normal'' keywords intended to simulate basic searches that a typical user might perform on the platform (see Table~\ref{tab:keywords}), briefly hovering over the first ten results (5--10 seconds each) without playing them in full. After this, the account proceeds with another 100-passive-video scroll. Subsequently, the user performs a second search using ``low-risk'' keywords, intending to simulate searches that are curious and potentially harmful but not explicitly so, again hovering over the initial ten results. Finally, the account scrolls through the last set of 100 videos, maintaining the 20-second watch duration per video. As before, non-English videos are skipped immediately. By examining these sequential steps, we can see how even searching for seemingly benign or ``low-risk'' terms can unintentionally increase exposure to problematic or harmful content.

\begin{table}
    \centering
    \caption{Keywords used for search-based scrolling.}
    \label{tab:keywords}
    \begin{tabular}{ll}
        \hline
        \textbf{Normal Keywords} & \textbf{Low-Risk Keywords} \\
        \hline
        football & Fake news \\
        make-up  & Free game codes \\
        games    & Buy followers \\
        happy    & prank call \\
        sad      & calories \\
                 & flirting \\
                 & depressed \\
                 & Hackers \\
                 & Conspiracy \\
                 & Challenge \\
                 & Fake ID \\
                 & swatting \\
                 & fasting \\
                 & grope \\
                 & unalive \\
        \hline
    \end{tabular}
\end{table}

\subsection{Account Creation}

To systematically compare content moderation outcomes across age groups and user behaviours, we created separate sets of accounts for each platform at two age brackets: 13 (minor) and 18 (adult). Four new accounts were generated for TikTok and YouTube: one minor and one adult for passive scrolling, and one minor and one adult for search-based scrolling. This design enables direct moderation efficacy assessments that account for user age and interaction styles.

On Instagram, the web-based interface primarily returns user profiles rather than content-based search results. This means that instead of surfacing individual posts, videos, or trending content in response to queries, Instagram emphasises user accounts matching the search terms. As a result, the ability to explore broader themes or topics via search is limited, especially for new or passive accounts. Consequently, we established only two Instagram accounts for each age group, which were limited to passive scrolling. In contrast, YouTube and TikTok supported the complete four-account model (passive and search-based, each for ages 13 and 18), as their search functionalities return diverse content types (e.g., videos, channels, hashtags), making them more conducive to studying the impact of searching for keywords. Every account was linked to a unique email address; however, the same email addresses could be reused across different platforms.

A particular challenge arose with YouTube’s requirements for minors under 16. During account creation, the platform automatically prompted parental supervision, requiring a parent or guardian to configure content restrictions. These options range from highly restrictive (only approved content) to nearly adult-like access. For this study, both child accounts were configured with the least restrictive parental settings. This approach was chosen to realistically simulate a scenario in which a minor sets up their account with minimal oversight, reflecting real-world behaviour \citep{icissp25} and demonstrating how these safeguards can be bypassed or loosely enforced. By doing so, we aimed to assess the actual exposure risks faced by underage users who may not have parents who actively engage with them. Table~\ref{tab:account_setup} summarises the final account configurations.

\begin{table}[tb]
    \centering
    \caption{Overview of account configurations across YouTube, Instagram and TikTok for 13-year-old and 18-year-old accounts.}
    \label{tab:account_setup}
    \begin{tabular}{c|c|l}
        \hline
        \textbf{Platform} & \textbf{Age} & \textbf{Method} \\
        \hline
        \multirow{4}{*}{TikTok} & \multirow{2}{*}{13 (Minor)} & Passive scroll \\
                                &                     & Search-based scroll \\
                                & \multirow{2}{*}{18 (Adult)} & Passive scroll \\
                                &                     & Search-based scroll \\
        \hline
        \multirow{4}{*}{YouTube} & \multirow{2}{*}{13 (Minor)} & Passive scroll \\
                                 &                     & Search-based scroll \\
                                 & \multirow{2}{*}{18 (Adult)} & Passive scroll \\
                                 &                     & Search-based scroll \\
        \hline
        \multirow{2}{*}{Instagram} & 13 (Minor)& Passive scroll \\
                                   & 18 (Adult) & Passive scroll \\
        \hline
    \end{tabular}
\end{table}

By consistently defining user ages and scrolling behaviours during account setup, we ensured reproducible results and minimised confounding variables across experiments.

\subsection{Labelling Methodology} \label{sec:labelling methods}

We adopted a multi-tiered labelling procedure, with particular attention to inter-annotator reliability and conflict resolution mechanisms.

Our primary classification process begins by tagging each video as \emph{Not Harmful} (video does not fall under any of the categories defined in the framework of harmful content) or \emph{Harmful} (video falls under any of the categories listed in the framework of harmful content defined earlier). Clips labelled \emph{Harmful} are then further assigned to one of the harmful content categories from our framework (see Methodology Section) and given a severity rating: \emph{Low}, \emph{Medium}, or \emph{High}.

\emph{Low} severity content typically includes material that engages with sensitive themes in a humorous or satirical manner. For instance, we encountered stand-up comedy clips joking about politics and identity in a self-aware, non-malicious tone. Similarly, several parody podcast segments mimicked controversial internet personalities for comedic effect, without promoting the views being mocked.

\emph{High} severity was used when videos contained explicit and severe forms of harm. In some cases, we observed individuals using a racial slur (i.e., the N-word) aggressively during rants, making blatantly sexist remarks about women’s roles in society, or expressing homophobic views without irony or critique. These videos demonstrated a clear intent to demean or provoke.

\emph{Medium} severity was assigned in cases where interpretation was mixed. For example, one clip featured a creator using inflammatory language in a comedic skit, but the satirical framing was not immediately obvious. While one annotator viewed it as a parody, another found the content indistinguishable from serious hate speech. In such cases, where the perceived intent or harm level varied significantly between annotators, we defaulted to a medium severity rating.

Table \ref{tab:examples} categorises some examples of videos based on their severity classification across a few categories.

\begin{table*}[tb]
    \centering
    \caption{Examples of Content by Severity and Category}
    \label{tab:examples}
    \begin{tabular}{>{\raggedright\arraybackslash}p{3.8cm} >{\raggedright\arraybackslash}p{2cm} >{\raggedright\arraybackslash}p{10.5cm}}
    \toprule
    \textbf{Category} & \textbf{Severity} & \textbf{Examples (as observed during annotation)} \\
    \midrule
    \multirow{2}{=}{Violent and Criminal Behaviour} 
    & Low & A stand-up comedy set where the comedian joked about political revolutions, using irony to highlight the absurdity of extremist views. \\
   
    & High & One video about how you can hit women if you transition into a woman. \\
    \midrule
    \multirow{2}{=}{Hate Speech and Hateful Behaviour} 
    & Low & A parody podcast in which the host mimicked a controversial online figure using exaggerated speech and tone, seemingly to mock rather than promote hateful ideology. \\
    & High & In one clip, a creator used the N-word repeatedly during an aggressive rant, alongside misogynistic and homophobic comments delivered without satire or critique. \\
    \midrule
    \multirow{2}{=}{Mental and Behavioural Health} 
    & Low & A video skit referenced a viral "breaking into dorm" circulating online, but presented them with a humorous tone, not clearly encouraging participation. \\
    & High & We came across a video where a person was drunk and was flaunting the fact that they were drunk while driving. \\
    \bottomrule
    \end{tabular}
\end{table*}

Initially, a single annotator (Labeller 1) evaluates each video for harm status, category, and severity. If Labeller 1 flags the video as Harmful, a second annotator (Labeller 2) reviews it to either confirm or dispute the label. If Labeller 2 disagrees, either by labelling it Not Harmful or by assigning a different category or severity, a third annotator (Labeller 3) is used for a final review.

For example, suppose Labeller 1 classifies a video as ``Harmful" under the category ``Violent Behaviour" with a Low severity. Labeller 2, upon review, disagrees and marks the video as \emph{Not Harmful}, triggering the need for Labeller 3’s judgment. If Labeller 3 agrees with either Labeller 1 or 2, either confirming the video is harmful and violent, or agreeing it is not harmful at all, the majority vote (2:1) is accepted as the final decision.

However, if Labeller 3 introduces a new perspective, for instance, labelling the content as \emph{Harmful} but under a different category, such as ``Mental and Behavioural Health", then there is no majority consensus. In such a case, a fourth annotator (Labeller 4) is brought in to make a final decision. 

This hierarchical process addresses several challenges inherent to content moderation. First, some videos span multiple categories (e.g., a self-harm video might also feature explicit or graphic themes). In such cases, the final label selected is the one agreed upon by at least two annotators. Second, severity can be subjective. For instance, mild hate speech might be classified as \emph{Low} severity by one annotator and \emph{High} severity by another; in these instances, we apply a rule that if there is a disagreement between two levels (e.g., Low vs. High), the rating defaults to Medium.

Two categories from our Harmful Content Framework, \emph{Privacy and Security} and \emph{Enforcement Actions}, were excluded from manual annotation. These categories address content beyond immediate view (e.g., personal data leaks, platform-level enforcement decisions) and thus could not be reliably observed through passive or minimal interactive engagement alone.

\subsection{Methodology Considerations}

Harm is a subjective and context-dependent construct, influenced by individuals’ cultural backgrounds, lived experiences, and levels of vulnerability \citep{context}. In addition, the assessment of content harm in the study must, by virtue of the nature of the research task, be carried out by adult researchers. This introduces a potential disconnect: adults can interpret tone, nuance, or severity differently from adolescents, especially in borderline or ambiguous cases \citep{smith2024youth, kumar2021designingtoxiccontentclassification}. Indeed, even across adolescents, the level of harm or perception of harm may differ in line with individual vulnerabilities, culture and other personal factors. Awareness of these discrepancies are important when evaluating harm thresholds and may influence how certain videos are classified and interpreted between age groups. However, while being aware of this subjectivity is important when interpreting results, manual labeling against a defined unified framework for harmful content helps to alleviate the impact of subjectivity, providing a published reference point of content definition and harm levels alongside the results.  
 
Some content flagged as “low-risk”, such as mental health or sexual health conveyed in a humorous or satirical way, tread  a fine line between harmful and informative. In certain contexts, these forms of expression could in theory increase relatability and engagement among youth, particularly for topics typically stigmatised or avoided in more formal messaging \citep{corrigan2014does, albers2022discussing}. While such content raises moderation raises concerns particularly when unconstrained by age, its presence in recommendation streams could also reflect how young audiences seek out accessible, peer-like discourse on sensitive issues in an adult-free environment.
 
 We deliberately restricted user interaction to an absolute minimum, avoiding actions such as liking, commenting, following, or rewatching videos, in order to observe the default content served by the platform algorithms without behavioural influence. This passive approach was chosen to simulate the experience of a newly created or minimally active user and to isolate how platforms push content based solely on the user's age.
 
 All non-English videos were skipped during the annotation phase to simplify evaluation and reduce linguistic ambiguity among annotators. While this choice enhanced consistency, it also excludes a substantial portion of global content and may have resulted in the omission of potentially harmful material presented in other languages.

\section{Analysis of the Results} \label{sec:Analysis_of_the_Results}

This section presents key findings on how age, interaction mode, and platform policies jointly shape users’ exposure to harmful content. Our analysis integrates several data points from the experiments including overall exposure rates to harmful content across all the categories defined - including severity ratings per video deemed as harmful, time to first harmful video for every account, differences in total harmful content exposure for the 13-year-old and the 18-year-old account, and differences in exposure to harmful content based on user behaviour (passive scrolling vs searching and scrolling), and reveals patterns that underscore the need for more robust age-specific moderation strategies.

\subsection{Age-Based Trends in Harmful Content Exposure}

\begin{figure}
\centerline{%
\includegraphics[width=0.9\textwidth]{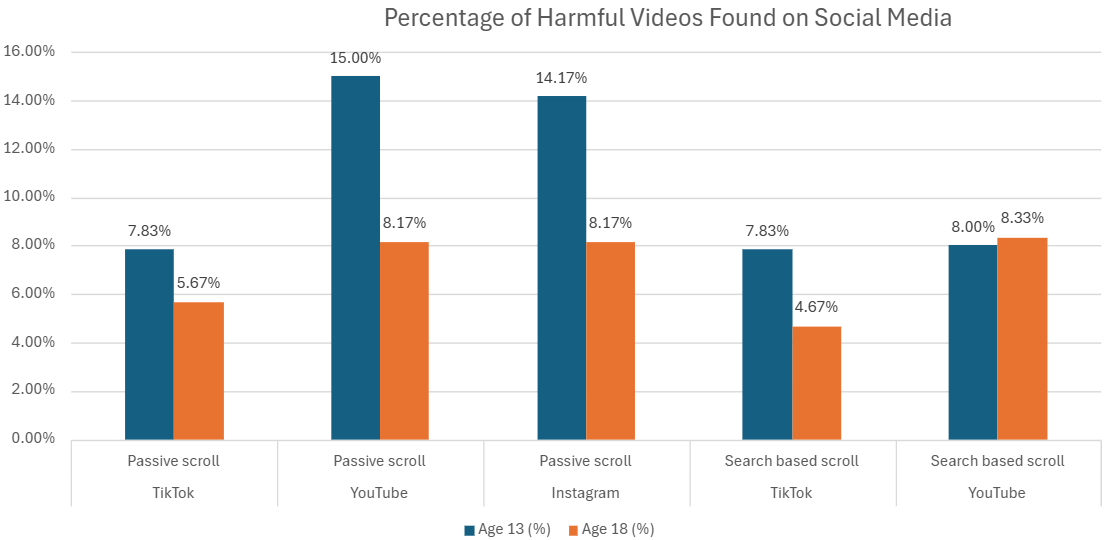}%
}%
\caption{Comparison of age-based harmful content trends.}
\label{fig:Trends}
\end{figure}

To assess whether platforms effectively adjust their moderation practices for younger audiences, we compared user accounts from 13-year-olds and 18-year-olds under identical conditions. As shown in Figure~\ref{fig:Trends}, \emph{13-year-old accounts generally received higher levels of harmful content across most platforms and interaction modes}. The only exception was YouTube in the search-based scenario, where both age groups faced nearly identical exposure rates.

A key finding is that harmful content for 13-year-olds spanned from 7.83\% to 15\%, while older users typically experienced lower exposure rates (4.67\% to 8.33\%). Even modest percentages for younger audiences are concerning, since no platform should recommend damaging material to minors. These results reinforce the importance of \emph{age-specific moderation protocols} designed to provide stronger safeguards for younger users.

\subsection{Impact of User Interaction Modes on Harmful Content Recommendations}

We then examined how different interaction styles affect recommendations of harmful content. Specifically, we compared \textit{passive scrolling}, which involves no active engagement (such as likes, searches, or comments), with \textit{search-based scrolling}, which includes targeted keyword searches alongside passive viewing. Section~\ref{User interaction modes} details these procedures.

\begin{figure}
 \centering
 \subfloat[]{{\includegraphics[width=0.65\textwidth]{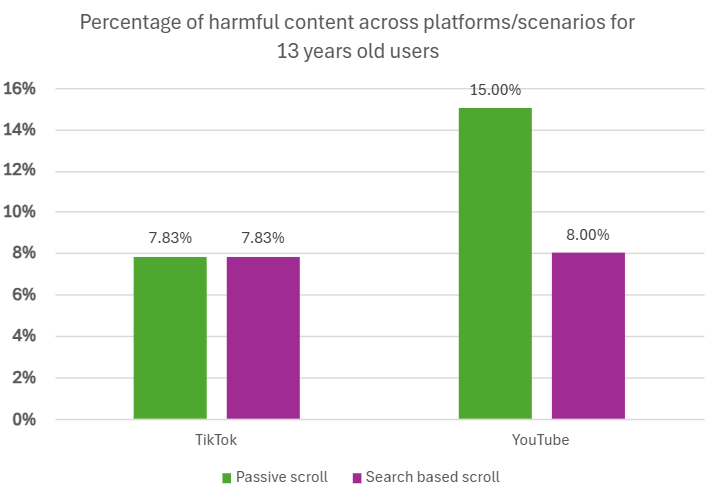}}\label{a-Comparison}}%
 \hspace{0.5cm}
 \subfloat[]{{\includegraphics[width=0.65\textwidth]{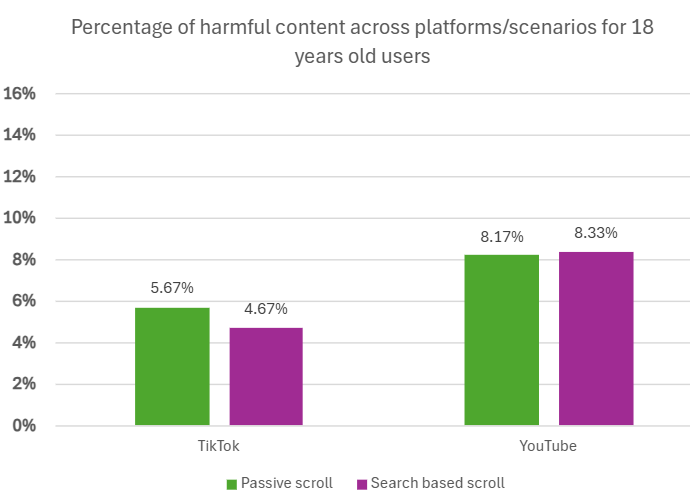}}\label{b-Comparison}}%
 \caption{Analysis of harmful content recommendations across platforms and scenarios for (a) 13-year-old users and (b) 18-year-old users.}
 \label{fig:Comparison}
 \end{figure}

Figure~\ref{fig:Comparison} shows the proportion of harmful content encountered by 13-year-old and 18-year-old users on YouTube and TikTok under both modes. For adults (Figure~\ref{b-Comparison}), TikTok exhibited a slight reduction in harmful content when users actively searched (from 5.67\% to 4.67\%), whereas YouTube’s rates remained essentially unchanged (8.17\% vs.\ 8.33\%).

For 13-year-olds (Figure~\ref{a-Comparison}), the exposure to harmful content on TikTok remained unchanged at 7.83\% when users conducted searches, whereas YouTube’s dropped significantly (from 15\% to 8\%). Together, these mixed outcomes suggest that \emph{search-based interactions} can either dampen or exacerbate exposure to harmful material, depending on the platform’s recommendation algorithms.

\subsection{Impact of Search Behaviour on Harmful Content Exposure}

\begin{figure}
    \centering
    \includegraphics[width=0.75\linewidth]{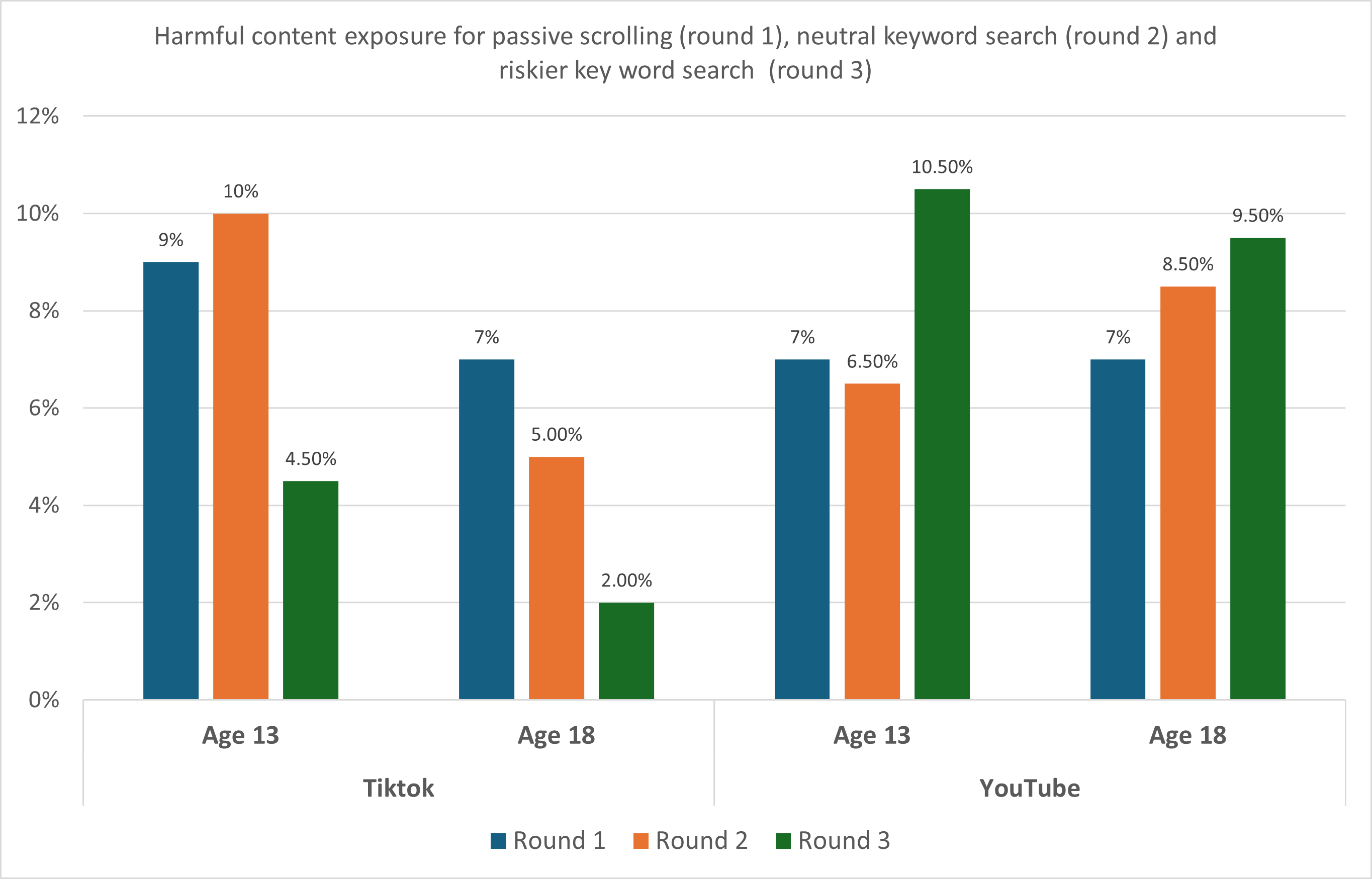}
    \caption{Impact of search behaviour on harmful content exposure.}
    \label{rounds}
\end{figure}

To further disentangle how specific keyword searches influence exposure, we tracked recommendations over three sequential “rounds” (Figure~\ref{rounds}). Each round involved scrolling through 100 videos: (1) purely passive scrolling; (2) searching with neutral keywords (e.g., “football”); and (3) searching with riskier keywords (e.g., “fasting”).

On TikTok, the amount of harmful content for younger users increased following the neutral search but declined by the final round. On YouTube, rates remained relatively steady, with a slight uptick for younger users after the low-risk keywords. This pattern implies that specific keyword searches may not necessarily push vulnerable users toward more harmful material, highlighting potential pitfalls in platform-based content filtering.

\subsection{Time to First Harmful Video Across Platforms}

\begin{figure}
    \centering
    \includegraphics[width=0.75\linewidth]{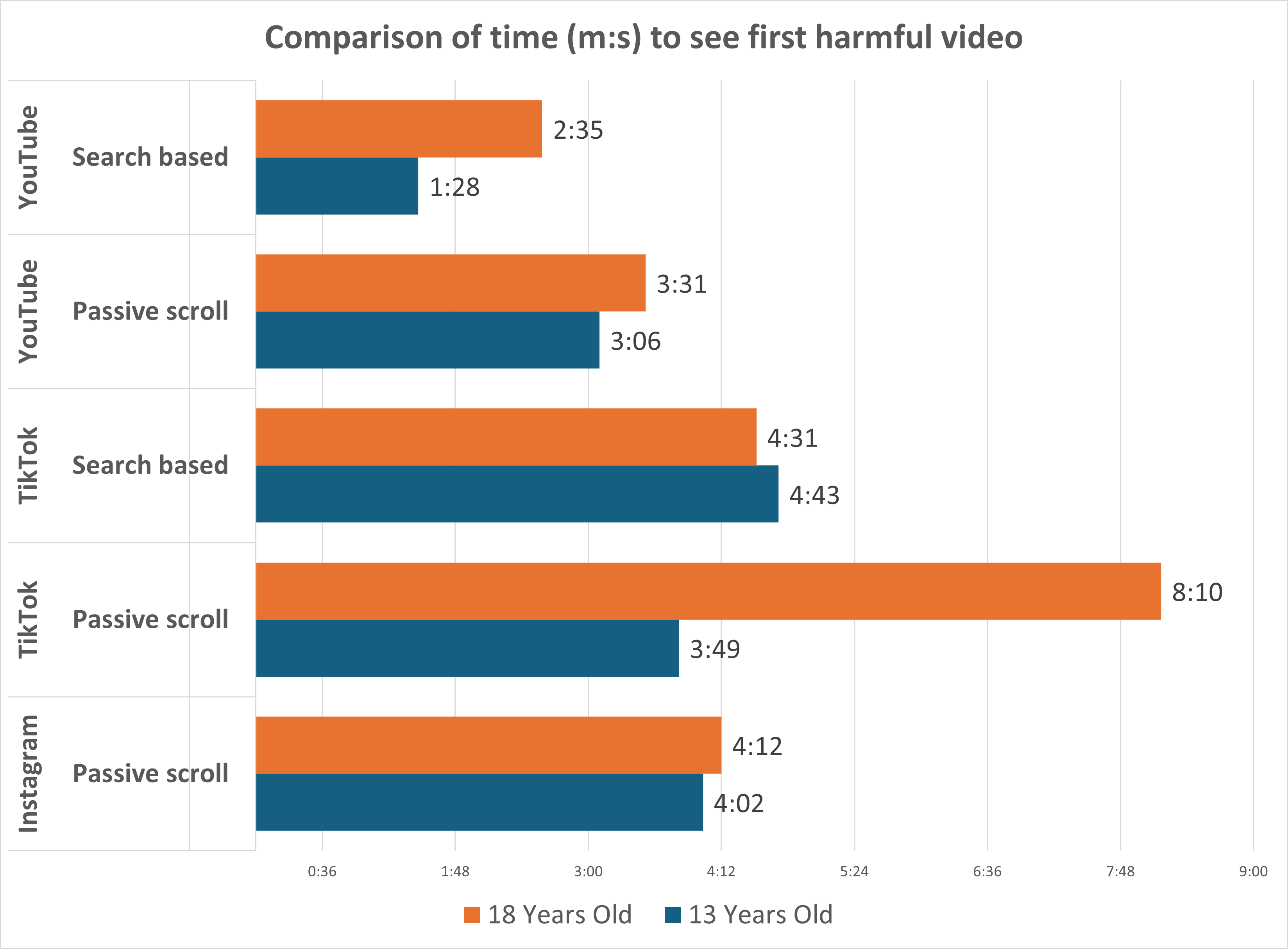}
    \caption{Time (\emph{minutes:seconds}) to the first harmful video across platforms and age groups.}
    \label{fig:Time}
\end{figure}

Another measure of moderation quality is how quickly new users encounter potentially harmful videos. Figure~\ref{fig:Time} illustrates the elapsed time before a user’s first exposure to harmful content, broken down by platform, age group, and interaction mode.



As shown in Figure~\ref{fig:Time}, younger users (13-year-olds) typically see harmful content faster than older users (18-year-olds). Under passive scrolling, YouTube (3:06~minutes) and TikTok (3:49~minutes) enable near-immediate exposure for minors. Search-based scrolling trends, however, differ by platform: on TikTok, it slightly \emph{increases} the time to first harmful clip (4:43~minutes vs.\ 3:49~minutes), whereas on YouTube, it \emph{reduces} it significantly (1:28~minutes vs.\ 3:06~minutes). Overall, children commonly encounter harmful content in under five minutes, compared to roughly nine minutes for adults, raising concerns about the current safeguards’ capacity to prevent early, potentially harmful exposure.

\subsection{Types and Severity Levels of Harmful Content Encountered}

\begin{figure}
\centering
\includegraphics[width=1\textwidth]{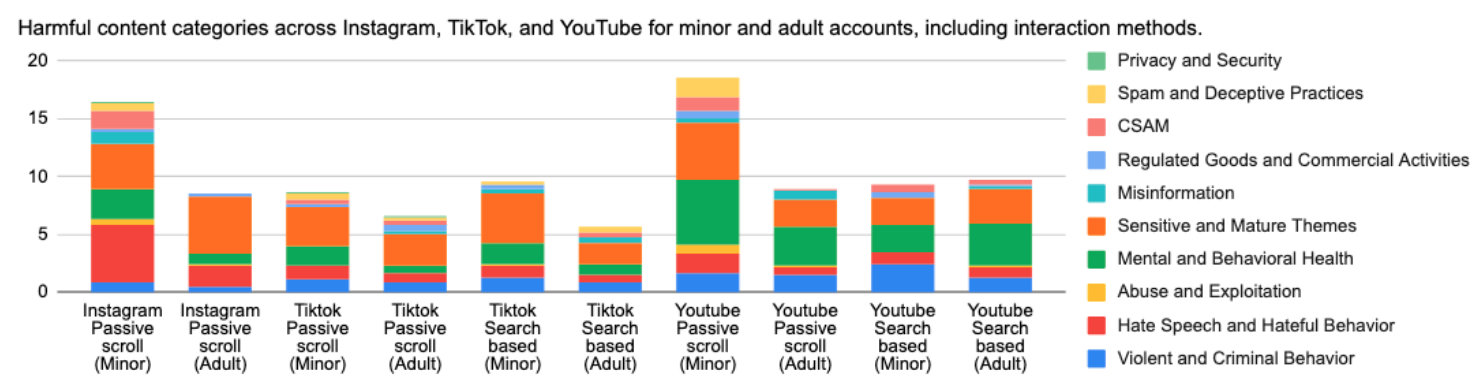}
\caption{Percentage Distribution of harmful content categories across different platforms and interaction methods for (a) 13-year-old users and (b) 18-year-old users.}
\label{ComparisonHarmfulCategories}
\end{figure}

We next examined the types of harmful content that users encountered most often, along with its severity. Figure~\ref{ComparisonHarmfulCategories} shows category-level distributions for 13- and 18-year-old user accounts on YouTube, TikTok, and Instagram, each under passive or search-based scrolling.

Across all user groups, \emph{Sensitive \& Mature Themes} consistently dominate recommended harmful content. In addition, 13-year-old Instagram users exhibit notably higher exposure (5\%) to \emph{Hate Speech and Hateful Behaviour}.

Harmful videos also vary in severity. Table~\ref{tab:harmful_content_distribution} employs a heatmap colour scheme to represent severity levels, where yellow denotes low severity, orange shades indicate medium severity, and red signifies high severity. The intensity of each colour varies according to the corresponding percentage, effectively illustrating both the severity and prevalence of harmful content categories across different scrolling methods. On TikTok, 13-year-olds encountered more low-severity \emph{Sensitive \& Mature Themes} under both passive (3\%) and search-based (4.17\%) scrolling. These younger users still faced more harmful content than older users, whose exposure rates declined across most categories.

\begin{table}
\centering
\caption{Distribution of Harmful Content Categories by (Platform \& Age), Method, and Severity (in \%).}
\label{tab:harmful_content_distribution}
\smallskip

\newcommand{\Lsev}{\cellcolor{yellow}Low}
\newcommand{\Msev}{\cellcolor{orange}Medium}
\newcommand{\Hsev}{\cellcolor{red}High}

\begin{adjustbox}{max width=\textwidth}
\begin{tabular}{
    >{\raggedright\arraybackslash}p{2cm}  
    p{3cm}                                
    c                                     
    S[table-format=1.2]  
    S[table-format=1.2]  
    S[table-format=1.2]  
    S[table-format=1.2]  
    S[table-format=1.2]  
    S[table-format=1.2]  
    S[table-format=1.2]  
    S[table-format=1.2]  
    S[table-format=1.2]  
    S[table-format=1.2]  
}

\rowcolor{blue!15}
\toprule
\multicolumn{2}{c}{\textbf{User Info}} &
\multicolumn{1}{c}{\textbf{\shortstack{Severity}}} &
\multicolumn{10}{c}{\textbf{Harmful Content Categories (\%)}} \\
\cmidrule(lr){1-2} \cmidrule(lr){3-13}

\rowcolor{blue!15}
\textbf{\shortstack{Platform\\ \& Age}} &
\textbf{\shortstack{Method\\(scroll/search)}} &
\textbf{} &
\multicolumn{1}{c}{\shortstack{\textbf{Viol.}\\ \textbf{\&Crim.}}} &
\multicolumn{1}{c}{\shortstack{\textbf{Hate}\\ \textbf{Speech}}} &
\multicolumn{1}{c}{\shortstack{\textbf{Abuse}\\ \textbf{\&Expl.}}} &
\multicolumn{1}{c}{\shortstack{\textbf{Mental}\\ \textbf{Health}}} &
\multicolumn{1}{c}{\shortstack{\textbf{Sens.}\\ \textbf{Themes}}} &
\multicolumn{1}{c}{\textbf{Misinfo.}} &
\multicolumn{1}{c}{\shortstack{\textbf{Reg.}\\ \textbf{Goods}}} &
\multicolumn{1}{c}{\textbf{CSAM}} &
\multicolumn{1}{c}{\textbf{Spam}} &
\multicolumn{1}{c}{\shortstack{\textbf{Privacy}\\ \textbf{\&Sec.}}} \\
\midrule

\multirow{6}{*}{\shortstack{TikTok\\13}} 
  & \multirow{3}{*}{Passive scroll}
    & \Lsev & \cellcolor{yellow!40}{0.83} & \cellcolor{yellow!40}1.17 & \cellcolor{yellow!20}0.00 &\cellcolor{yellow!40} 1.33 &\cellcolor{yellow!60}{3.00} & \cellcolor{yellow!20} 0.00 &\cellcolor{yellow!40} 0.33 & \cellcolor{yellow!40}0.17 &\cellcolor{yellow!40} 0.50 & \cellcolor{yellow!40}0.17 \\
    
  & & \Msev & \cellcolor{orange!20}0.00 & \cellcolor{orange!20}0.00 & \cellcolor{orange!20}0.00 & \cellcolor{orange!40}0.33 & \cellcolor{orange!40}0.17 & \cellcolor{orange!20}0.00 & \cellcolor{orange!20}0.00 &\cellcolor{orange!20} 0.00 & \cellcolor{orange!20}0.00 & \cellcolor{orange!20}0.00 \\
  & & \Hsev & \cellcolor{red!40}0.33 & \cellcolor{red!20}0.00 & \cellcolor{red!20}0.00 & \cellcolor{red!20}0.00 & \cellcolor{red!40}0.17 & \cellcolor{red!20}0.00 & \cellcolor{red!20}0.00 & \cellcolor{red!40}0.17 &\cellcolor{red!20} 0.00 & \cellcolor{red!20}0.00 \\
 \cline{2-13}\cline{2-13}

  & \multirow{3}{*}{Search based scroll}
    & \Lsev &\cellcolor{yellow!40} 1.33 & \cellcolor{yellow!40}1.00 & \cellcolor{yellow!40}0.17 & \cellcolor{yellow!40}1.67 & \cellcolor{yellow!80}{4.17} & \cellcolor{yellow!40}0.50 & \cellcolor{yellow!40}0.33 &  \cellcolor{yellow!20} 0.00 & \cellcolor{yellow!40}0.33 &  \cellcolor{yellow!20} 0.00 \\
  & & \Msev & \cellcolor{orange!20}0.00 & \cellcolor{orange!20}0.00 & \cellcolor{orange!20}0.00 & \cellcolor{orange!40}0.17 & \cellcolor{orange!20}0.00 & \cellcolor{orange!20}0.00 & \cellcolor{orange!20}0.00 & \cellcolor{orange!20}0.00 & \cellcolor{orange!20}0.00 & \cellcolor{orange!20}0.00 \\
  & & \Hsev &  \cellcolor{red!20}0.00 &  \cellcolor{red!20}0.00 &  \cellcolor{red!20}0.00 &  \cellcolor{red!20}0.00 &  \cellcolor{red!20}0.00 &  \cellcolor{red!20}0.00 &  \cellcolor{red!20}0.00 &  \cellcolor{red!20}0.00 &  \cellcolor{red!20}0.00 &  \cellcolor{red!20}0.00 \\
\midrule

\multirow{6}{*}{\shortstack{TikTok\\18}}
  & \multirow{3}{*}{Passive scroll}
    & \Lsev & \cellcolor{yellow!40}0.67 & \cellcolor{yellow!40}0.50 &  \cellcolor{yellow!20}0.00 &\cellcolor{yellow!40} 0.67 & \cellcolor{yellow!60}{2.17} & \cellcolor{yellow!40}0.33 & \cellcolor{yellow!40}0.50 & \cellcolor{yellow!40}0.33 & \cellcolor{yellow!40}0.33 & \cellcolor{yellow!40}0.17 \\
    
  & & \Msev &  \cellcolor{orange!40}0.17 & \cellcolor{orange!40}0.33 &  \cellcolor{orange!20}0.00 &\cellcolor{orange!20} 0.00 & \cellcolor{orange!40}0.50 & \cellcolor{orange!20}0.00 &\cellcolor{orange!20} 0.00 & \cellcolor{orange!20}0.00 & \cellcolor{orange!20}0.00 & \cellcolor{orange!20}0.00 \\
  
  & & \Hsev & \cellcolor{red!20}0.00 &  \cellcolor{red!20}0.00 &  \cellcolor{red!20}0.00 &  \cellcolor{red!20}0.00 &  \cellcolor{red!20}0.00 &  \cellcolor{red!20}0.00 &  \cellcolor{red!20}0.00 &  \cellcolor{red!20}0.00 &  \cellcolor{red!20}0.00 & \cellcolor{red!20} 0.00 \\
\cline{2-13}
  & \multirow{3}{*}{Search based scroll}
    & \Lsev & \cellcolor{yellow!40}0.83 & \cellcolor{yellow!40}0.67 &  \cellcolor{yellow!20}0.00 & \cellcolor{yellow!40}1.00 & \cellcolor{yellow!40}1.83 & \cellcolor{yellow!40}0.50 & \cellcolor{yellow!20} 0.00 & \cellcolor{yellow!40}0.33 &\cellcolor{yellow!40} 0.50 &  \cellcolor{yellow!20}0.00 \\
  & & \Msev & \cellcolor{orange!20}0.00 & \cellcolor{orange!20}0.00 & \cellcolor{orange!20}0.00 & \cellcolor{orange!20}0.00 & \cellcolor{orange!20}0.00 & \cellcolor{orange!20}0.00 & \cellcolor{orange!20}0.00 & \cellcolor{orange!20}0.00 & \cellcolor{orange!20}0.00 & \cellcolor{orange!20}0.00 \\
  
  & & \Hsev &  \cellcolor{red!20}0.00 &  \cellcolor{red!20}0.00 &  \cellcolor{red!20}0.00 &  \cellcolor{red!20}0.00 &  \cellcolor{red!20}0.00 &  \cellcolor{red!20}0.00 &  \cellcolor{red!20}0.00 &  \cellcolor{red!20}0.00 &  \cellcolor{red!20}0.00 &  \cellcolor{red!20}0.00 \\
\midrule
\midrule

\multirow{6}{*}{\shortstack{YouTube\\13}}
  & \multirow{3}{*}{Passive scroll}
    & \Lsev & \cellcolor{yellow!40}1.67 & \cellcolor{yellow!40}1.67 & \cellcolor{yellow!40}0.83 & \cellcolor{yellow!80}{5.17} &\cellcolor{yellow!80}{5.00} & \cellcolor{yellow!40}0.33 &\cellcolor{yellow!40} 0.50 &\cellcolor{yellow!40} 1.17 &\cellcolor{yellow!40} 1.67 &\cellcolor{yellow!20} 0.00 \\
    
  & & \Msev &\cellcolor{orange!20} 0.00 & \cellcolor{orange!20}0.00 & \cellcolor{orange!20}0.00 & \cellcolor{orange!40}0.33 & \cellcolor{orange!20}0.00 & \cellcolor{orange!20}0.00 & \cellcolor{orange!40}0.17 & \cellcolor{orange!20}0.00 & \cellcolor{orange!20}0.00 & \cellcolor{orange!20}0.00 \\
  
  & & \Hsev & \cellcolor{red!20}0.00 & \cellcolor{red!20}0.00 & \cellcolor{red!20}0.00 & \cellcolor{red!20}0.00 & \cellcolor{red!20}0.00 & \cellcolor{red!20}0.00 & \cellcolor{red!20}0.00 & \cellcolor{red!20}0.00 & \cellcolor{red!20}0.00 & \cellcolor{red!20}0.00 \\
\cline{2-13}
  & \multirow{3}{*}{Search based scroll}
    & \Lsev & \cellcolor{yellow!60}{2.33} &\cellcolor{yellow!40} 1.00 & \cellcolor{yellow!20}0.00 & \cellcolor{yellow!40}1.50 & \cellcolor{yellow!60}{2.17} & \cellcolor{yellow!20}0.00 & \cellcolor{yellow!40}0.33 &\cellcolor{yellow!40} 0.67 & \cellcolor{yellow!20}0.00 & \cellcolor{yellow!20}0.00 \\
    
  & & \Msev & \cellcolor{orange!40}0.17 & \cellcolor{orange!20}0.00 & \cellcolor{orange!20}0.00 & \cellcolor{orange!40}0.83 & \cellcolor{orange!40}0.17 & \cellcolor{orange!20}0.00 & \cellcolor{orange!20}0.00 & \cellcolor{orange!20}0.00 & \cellcolor{orange!20}0.00 & \cellcolor{orange!20}0.00 \\
  
  & & \Hsev & \cellcolor{red!20}0.00 &\cellcolor{red!20} 0.00 & \cellcolor{red!20}0.00 & \cellcolor{red!20}0.00 & \cellcolor{red!20}0.00 &\cellcolor{red!20} 0.00 & \cellcolor{red!40}0.17 & \cellcolor{red!20}0.00 & \cellcolor{red!20}0.00 & \cellcolor{red!20}0.00 \\
\midrule

\multirow{6}{*}{\shortstack{YouTube\\18}}
  & \multirow{3}{*}{Passive scroll}
    & \Lsev & \cellcolor{yellow!40}1.33 & \cellcolor{yellow!40}0.67 & \cellcolor{yellow!40}0.17 & \cellcolor{yellow!60}{3.00} & \cellcolor{yellow!60}{2.33} & \cellcolor{yellow!40}0.83 & \cellcolor{yellow!20}0.00 & \cellcolor{yellow!40}0.17 & \cellcolor{yellow!20}0.00 & \cellcolor{yellow!20}0.00 \\
    
  & & \Msev & \cellcolor{orange!20}0.00 & \cellcolor{orange!20}0.00 & \cellcolor{orange!20}0.00 & \cellcolor{orange!40}0.17 & \cellcolor{orange!20}0.00 & \cellcolor{orange!20}0.00 & \cellcolor{orange!20}0.00 & \cellcolor{orange!20}0.00 & \cellcolor{orange!20}0.00 &\cellcolor{orange!20} 0.00 \\
  & & \Hsev & \cellcolor{red!40}0.17 & \cellcolor{red!20}0.00 & \cellcolor{red!20}0.00 & \cellcolor{red!40}0.17 & \cellcolor{red!20}0.00 & \cellcolor{red!20}0.00 & \cellcolor{red!20}0.00 & \cellcolor{red!20}0.00 & \cellcolor{red!20}0.00 & \cellcolor{red!20}0.00 \\
\cline{2-13}
  & \multirow{3}{*}{Search based scroll}
    & \Lsev & \cellcolor{yellow!40}1.17 & \cellcolor{yellow!40}0.83 & \cellcolor{yellow!40}0.17 & \cellcolor{yellow!60}{3.33} & \cellcolor{yellow!60}{3.00} & \cellcolor{yellow!40}0.17 & \cellcolor{yellow!40}0.17 & \cellcolor{yellow!40}0.17 & \cellcolor{yellow!20}0.00 & \cellcolor{yellow!20}0.00 \\
    
  & & \Msev & \cellcolor{orange!40}0.17 & \cellcolor{orange!20}0.00 & \cellcolor{orange!20}0.00 & \cellcolor{orange!40}0.33 & \cellcolor{orange!20}0.00 & \cellcolor{orange!20}0.00 & \cellcolor{orange!20}0.00 & \cellcolor{orange!40}0.17 & \cellcolor{orange!20}0.00 & \cellcolor{orange!20}0.00 \\
  & & \Hsev & \cellcolor{red!20}0.00 & \cellcolor{red!20}0.00 & \cellcolor{red!20}0.00 & \cellcolor{red!20}0.00 & \cellcolor{red!20}0.00 &\cellcolor{red!20} 0.00 & \cellcolor{red!20}0.00 & \cellcolor{red!20}0.00 & \cellcolor{red!20}0.00 &\cellcolor{red!20} 0.00 \\
\midrule
\midrule

\multirow{3}{*}{\shortstack{Instagram\\13}}
  & \multirow{3}{*}{Passive scroll}
    & \Lsev &\cellcolor{yellow!40} 0.83 & \cellcolor{yellow!80}{5.00} & \cellcolor{yellow!40}0.33 & \cellcolor{yellow!60}{2.50} & \cellcolor{yellow!60}{3.50} &\cellcolor{yellow!40} 1.00 & \cellcolor{yellow!40}0.33 &\cellcolor{yellow!40} 1.33 & \cellcolor{yellow!40}0.67 &\cellcolor{yellow!40} 0.17 \\
  & & \Msev & \cellcolor{orange!20}0.00 & \cellcolor{orange!20}0.00 & \cellcolor{orange!40}0.17 & \cellcolor{orange!40}0.17 & \cellcolor{orange!40}0.33 & \cellcolor{orange!20}0.00 & \cellcolor{orange!20}0.00 & \cellcolor{orange!40}0.17 & \cellcolor{orange!20}0.00 & \cellcolor{orange!20}0.00 \\
  & & \Hsev & \cellcolor{red!20}0.00 & \cellcolor{red!20}0.00 & \cellcolor{red!20}0.00 & \cellcolor{red!20}0.00 & \cellcolor{red!20}0.00 & \cellcolor{red!20}0.00 & \cellcolor{red!20}0.00 & \cellcolor{red!20}0.00 & \cellcolor{red!20}0.00 & \cellcolor{red!20}0.00 \\
\midrule

\multirow{3}{*}{\shortstack{Instagram\\18}}
  & \multirow{3}{*}{Passive scroll}
    & \Lsev &\cellcolor{yellow!40} 0.33 &\cellcolor{yellow!40} 1.33 & \cellcolor{yellow!40}0.17 & \cellcolor{yellow!40}0.67 & \cellcolor{yellow!80}{4.67} & \cellcolor{yellow!20}0.00 &\cellcolor{yellow!40} 0.17 & \cellcolor{yellow!20}0.00 & \cellcolor{yellow!20}0.00 & \cellcolor{yellow!20}0.00 \\
  & & \Msev & \cellcolor{orange!40}0.17 & \cellcolor{orange!40}0.50 & \cellcolor{orange!20}0.00 & \cellcolor{orange!40}0.17 & \cellcolor{orange!40}0.17 & \cellcolor{orange!20}0.00 & \cellcolor{orange!20}0.00 & \cellcolor{orange!20}0.00 & \cellcolor{orange!20}0.00 & \cellcolor{orange!20}0.00 \\
  & & \Hsev & \cellcolor{red!20}0.00 & \cellcolor{red!20}0.00 & \cellcolor{red!20}0.00 & \cellcolor{red!20}0.00 & \cellcolor{red!40}0.17 & \cellcolor{red!20}0.00 & \cellcolor{red!20}0.00 & \cellcolor{red!20}0.00 & \cellcolor{red!20}0.00 & \cellcolor{red!20}0.00 \\
\bottomrule

\end{tabular}
\end{adjustbox}
\end{table}

YouTube exhibited a larger overall share of harmful content, particularly among 13-year-olds. These users frequently encountered low-severity mental health–related content (5.17\% via passive scrolling) and other sensitive themes (5\%). Although active searching reduced exposure to some mental health–related harms, 18-year-olds still faced a nontrivial amount of such material (3.33\% via search-based scrolling).

Instagram presented a different pattern: 13-year-olds encountered a relatively high percentage of low-severity hate speech (5\%), sensitive themes (3.5\%), and mental health–related content (2.5\%). For 18-year-olds on Instagram, overall exposure to harmful content was lower, except for sensitive themes, which rose to 4.67\%.

Taken together, \textbf{low-severity harmful content} is the most prevalent form encountered across all platforms, categories, interaction types, and age brackets, suggesting a potential for normalisation over time. Repeated exposure, even to less extreme content, can desensitize users, particularly minors, to harmful speech and imagery \citep{desentitised}.

\subsection{Discussion and Key Findings} \label{Discussion of the Results}

Our findings indicate that \textbf{13-year-olds experience higher and faster exposure to harmful videos} compared to 18-year-olds in the majority of scenarios. For minors, exposure rates range from 7.83\% to 15\%; for adults, they average between 4.67\% and 8.33\%. Such disparities underscore the urgent need for \emph{age-specific videos moderation}: allowing minors to encounter harmful videos—particularly within minutes of use—is problematic from both ethical and safety standpoints.

One possible explanation for this pattern is that social media algorithms prioritise engagement, often recommending content that maximises watch time and interaction. Younger users may be more likely to engage with extreme videos, even unintentionally, which could lead to recommendation systems pushing increasingly harmful videos. This raises critical questions about whether current moderation strategies effectively safeguard younger audiences.

Interaction modes also impact exposure, though in platform-specific ways. On YouTube, \emph{search-based scrolling} significantly lowered harmful videos for minors compared to passive scrolling. Still, on TikTok, the extent of harmful recommendations for children is the same for both interaction modes. Moreover, risky keyword searches on YouTube amplified exposure, rising from 7\% to 10.5\% across sequential rounds.

Beyond mere percentage rates, children also encounter harm more \emph{quickly} (within five minutes) than adults do (under nine minutes). This finding raises serious questions about the real-time effectiveness of current safeguards in limiting minors’ early exposure. 

The dominant category of harmful videos for both age groups is \emph{Sensitive \& Mature Themes} such as violence, shocking content like car crashes, accidents, and some sexually suggestive content. However, 13-year-olds on Instagram also face a higher incidence of \emph{Hate Speech and Hateful Behaviour} like racial abuse and misogyny. Low-severity forms of harm, ranging from 0.17\% to 5\% across platforms, may appear minor yet pose cumulative risks of desensitisation and normalisation.

Overall, the data suggests that recommendation systems sometimes amplify harmful videos rather than suppressing them, exposing minors to significant risk. This highlights the need for more robust and transparent video moderation methods to effectively reduce such exposure.

\subsection{Implications for Content Moderation}

Based on our results, we propose the following recommendations to enhance user safety, particularly for minors:

\begin{itemize}
    \item \textbf{Strengthening Age-Specific Moderation:}  
    13-year-olds consistently experience more harmful material and at faster rates than older users. Platforms must adopt stricter filtering mechanisms tailored to younger audiences, mitigating exposure within the first few minutes of use.

    \item \textbf{Addressing Gaps in Enforcement:}
    Despite formal policies, \emph{Sensitive \& Mature Themes} and \emph{Hate Speech} remain pervasive. Platforms should refine their definitions of harmful content, increase consistency in enforcement, and update algorithms to respond more effectively to problematic material.

    \item \textbf{Mitigating Normalization of Low-Severity Harm:}
    Repetitive exposure to lower-level harm (e.g., insensitive jokes, mild stereotypes) can desensitise users over time. Moderation practices should prioritise not only extreme or violent incidents but also contain the routine circulation of ostensibly ‘‘minor’’ harmful content.

\end{itemize}

By integrating these improvements, social media platforms can align everyday recommendation algorithms with their expressed commitments to user safety. This alignment is particularly urgent for minors, given their heightened vulnerability and rapid exposure timelines.

\section{Conclusion} \label{sec:Conclusion}

This study presents an in-depth evaluation of the extent to which leading video-sharing social media platforms \emph{protect} young users from harmful content, examining both \emph{passive} and \emph{search-based} engagements on TikTok, YouTube, and Instagram. A comparison between 13-year-old and 18-year-old user accounts shows that minors face disproportionately higher levels of harmful videos, spanning 7--15\% versus 4--8\% for adults. In several cases, harmful videos were recommended to children's accounts within just three or four minutes, highlighting not only a greater volume but also a faster pace of exposure.

A severity-focused analysis reveals that \emph{low-severity} harmful videos are particularly widespread, potentially normalising harmful material and fostering desensitisation over time. Moreover, our findings expose inconsistencies in content definitions: some platforms explicitly list categories such as ``Privacy Violations'' or ``Dangerous Challenges,'' while others lack clarity, creating enforcement gaps and leaving younger users susceptible to incompletely regulated content.

Age-specific filtering and better alignment between platform policies and content moderation of video recommendation algorithms are, therefore, essential steps to reduce youth exposure to harmful material. Instead of prioritising engagement metrics, platforms must hold minors’ feeds to more stringent moderation standards. \emph{Stronger regulatory oversight} is also needed, driving platforms to implement uniform protections for minors and ensuring consistent enforcement of community guidelines.

The findings of this study highlight the pressing issue of minors' exposure to harmful and influential content on video-sharing platforms. Several avenues for future research may help build on this work. One potential direction involves repeating the study later to evaluate whether social media platforms have enhanced the effectiveness of their algorithms and moderation systems in safeguarding teenage users. Another promising area is the development and evaluation of educational interventions, such as game-based tools, to support children and their parents or guardians in recognising early signs of online grooming. Additionally, future research could explore the feasibility of device-level solutions designed to filter harmful content before it is displayed to minors.

Further investigation is also warranted into the subjectivity of content classification, particularly about low-level harm. This study relied on adult researchers to evaluate content, yet there may be notable differences in how adults and minors perceive and interpret such material. Content that adults perceive as low severity may be interpreted by younger users as highly distressing or threatening, due to their limited life experience or emotional maturity. Involving young people directly, through surveys, interviews, or focus groups, can help close the gap between how adults and minors understand harmful content. These methods allow young users to share how they feel about different types of content, making classification systems more accurate and better suited to their experiences.

It may also be valuable to examine the role of low-risk content that uses humour or satire to address sensitive topics such as mental health or sexual health. While such content may raise moderation concerns, it may also facilitate engagement and understanding among youth, particularly when formal messaging is less effective. Finally, the hypothesis that repeated exposure to low-severity harmful content contributes to desensitisation remains an important, though underexplored, research question. Empirical work in this area could help clarify the long-term impacts of such content on young audiences. Collectively, these directions offer opportunities to support a safer and more balanced digital environment for children and adolescents.

Additionally, future work could expand on the current methodology by incorporating a wider range of user behaviours, including engagement with specific content, to better understand how recommendation algorithms adapt and escalate exposure based on interaction as well as including multilingual content, ideally with the support of annotators fluent in relevant languages, to ensure broader cultural and linguistic representation in harm detection.

\section* {\uppercase{Acknowledgements}}

This paper is part of the N-Light project funded by the Safe Online Initiative of End Violence and the Tech Coalition through the Tech Coalition Safe Online Research Fund (Grant number: 21-EVAC-0008-Technological University Dublin).

\bibliographystyle{unsrt}  
\bibliography{references}  






\end{document}